# *Fast Exact Method for Solving the Travelling Salesman Problem*

**Vadim Yatsenko***

Nowadays Travelling Salesman Problem (TSP) is considered as NP-hard one. TSP exact solution of polynomial complexity is presented below.

TSP may be stated as follows [1-3]: let assume that $P^N = \{p_1, p_2, ..., p_N\}$ points set and $d(p_i, p_j)$ matrix of distances between these ones are determined. The problem is to determine a closed route $\overline{R}^N = \{r_1, r_2, ..., r_N\}$, $r_i \in P^N$, $i = 1,...,N$ that meets $D(\overline{R}^N) = \sum_{i=1}^{N, r_{N+1}=r_1} d(r_i, r_{i+1}) \to \min$ condition.

Let us suppose that somewise we succeed in finding $\overline{R}^N$ optimal route and consider $R^{N-1}(r_i) = \overline{R}^N - r_i$, $i = 1,...,N$ sub-routes set.

<u>*Lemma*</u> $\overline{R}^{N-1} = R^{N-1}(r_i) : D(R^{N-1}(r_i)) = \min_{j=1,...,N} D(R^{N-1}(r_j))$ route is the optimal one for $P^{N-1} = P^N - r_i$ points set.

Let apply this cutting procedure for $\overline{R}^{N-1} \to \overline{R}^{N-2} \to \overline{R}^{N-3}$... sub-routes. One can readily see that there will be progressive increasing in the disturbances of sub-routes length, i.e. $| D(\overline{R}^{k+1}) - D(\overline{R}^k) | \le | D(\overline{R}^k) - D(\overline{R}^{k-1}) |$, $k = N-2, N-3,...$ As result of application these transformations sequence, we should obtain $\overline{R}^2 = \{r_1, r_2\}$, $r_1, r_2 \in P^N$ sub-route, points of which must satisfy $d(r_1, r_2) = \max_{p_1, p_2 \in P^N} d(p_1, p_2)$ condition.

Supposing that above written cutting procedure could be inverted then the following recurrent adding algorithm for searching $\overline{R}^N$ optimal route could be gained.

1. Look for $p_i, p_j \in P^N$ points that satisfy $d(p_k, p_l) = \max_{p_i, p_j \in P^N} d(p_i, p_j)$ condition. Points $r_1 = p_k$, $r_2 = p_l$ will constitute $\overline{R}^2 = \{r_1, r_2\}$ initial route. Route length is given by $D(\overline{R}^2) = 2d(r_1, r_2)$
2. From the $P^N - \overline{R}^2$ points set, which are not included in $\overline{R}^2$ route, let be selected $p_i$ point that satisfies the $D(\overline{R}^2 + p_i) = \max_{p_j \in P^n - \overline{R}^2} (d(r_1, p_j) + d(r_2, p_j) + d(r_1, r_2))$ condition. It is assumed that $r_3 = p_i$, $\overline{R}^3 = \overline{R}^2 + r_3$. Route length is determined as $D(\overline{R}^3) = d(r_1, r_2) + d(r_2, r_3) + d(r_3, r_1)$. Note: selected $p_i$ point may be included in following ways $\overline{R}^3 = \{r_1, r_2, r_3\}$ or $\overline{R}^3 = \{r_1, r_3, r_2\}$; picked manner will determine the direction of optimal route - clockwise or counter-clockwise. This fact is not significant for considered case of symmetric TSP, but this one will be important for case of asymmetric TSP.
3. Further, for all $\overline{R}^i$, $i = 3,...,N-1$ routes, the following algorithm is to be applied iteratively. For each route edge $\{r_k, r_{k+1}\}$, $k = 1,...,i$, $r_{N+1} = r_1$ we have to select $p_j \in P^N - \overline{R}^i$ point that

* Yuzhnoye State Design Office. e-mail: Vadim_Yatsenko@ua.fm

provides minimization the disturbance of $D(\overline{R}^i)$ route length, i.e. $\Delta(r_k, p_j) = \min_{p_l \in P^N - \overline{R}^i}(d(p_l, r_k) + d(p_l, r_{k+1}) - d(r_k, r_{k+1}))$. From $\Delta(r_k, p_j)$, $k = 1,...,i$ disturbances set to be selected $\{r_m, r_{m+1}\}$ edge and $p_l$ point that meet $\Delta(r_m, p_l) = \max_{r_k \in \overline{R}^i} \Delta(r_k, p_j)$. Point $p_l$ have to be included in $\overline{R}^{i+1}$ route between $r_m$ and $r_{m+1}$ points, i.e. $r_{m+1} = p_l$ and $r_{s+1} = r_s$, $s = m+1, m+2,...$ Route length is given by $D(\overline{R}^{i+1}) = D(\overline{R}^i) + \Delta(r_m, p_l)$

Above method may be expressed by following recurrent relation $D(\overline{R}^{i+1}) = D(\overline{R}^i) + \max_{r_k \in \overline{R}^i} \min_{p_j \in P^N - \overline{R}^i} \Delta(r_k, p_j)$. Computational complexity of this algorithm may be evaluated by number of $\Delta(r_k, p_j)$ computations, which is given by $K \sim \sum_{k}^{N-1} k(N-k) \sim O^*(N^3)$.

One can readily see that in some cases the proposed procedure will not approach to $\overline{R}^N$ optimal route: there are possible situations, when computing of $\overline{R}^{i+1}$ intermediate routes (including $\overline{R}^2$ route) will provide a number of $\Delta(r_k, p_j)$ solutions that will satisfy the optimality condition, i.e. $\max_{r_k \in \overline{R}^i} \min_{p_j \in P^N - \overline{R}^i} \Delta(r_k, p_j) \to \{r_{m1}, p_{l1}\}, \{r_{m2}, p_{l2}\},...$ (for $\overline{R}^2$ intermediate route it may be the case of $\max_{p_i, p_j \in P^N} d(p_i, p_j) \to \{p_{k1}, p_{l1}\}, \{p_{k2}, p_{l2}\},...$). Under such condition, a necessity to search along all possible variants of $\overline{R}_j^{i+1}$ intermediate sub-routes (here $j$ – number of variant) could occur. Such a situation may emerge during searching along $\overline{R}_j^{i+1}$ intermediate sub-routes again, i.e. it is possible the appearance of some tree structure of sub-routes. In general, a conclusion about optimality of one variant or another must be reached after computing all $\overline{R}_j^N$ sub-routes. However, taking into account the logic of described above cutting procedure, the following heuristics should be true – if it will be determined that $\overline{R}_k^l : D(\overline{R}_k^l) = \max_n(D(\overline{R}_n^l))$, then this variant should bring to optimal route, and other sub-routes may be removed from the consideration. From all above follows that effectiveness of proposed method essentially decreased when there exist some $P^K \in P^N$ points sub-set, which possess some symmetry properties (common-type, local etc).

Computer modeling was carried out with the purpose of verifying an effectiveness of proposed method. Program code implemented "pure" algorithm without consideration of possible branching variants. $P^N$ points set was generated with application of random values generator, $d(p_i, p_j)$ matrix of distances was calculated, and then the searching algorithm itself has been run. Having no independent criteria for estimation the optimality of computed routes, it was assessed via visual inspection of the obtained results. The route has been considered as optimal one in case no loops were detected. Three basic options were considered: $N$=500, $N$=1000 and $N$=2000. Fig. 1, 2 and 3 present examples of routes those were determined as optimal. Modeling had included several tens of searching program, at that $P^N$ points set was generated anew. No loops have been detected any loops in the results of $N$=500 variant; runs of $N$=1000 variant have brought a loop occurrence in almost one-half cases; and, as of $N$=2000 variant, the loops appeared in about 4/5 cases of modeling outcomes. This fact may be plausible explained in a following way: applied algorithm of generation points set on some discrete grid had provided this one that for $N$=2000 option often offer some symmetry properties, and in turn confirms above written reasoning to the effect that in some cases "pure" method may approach to some sub-optimal route.

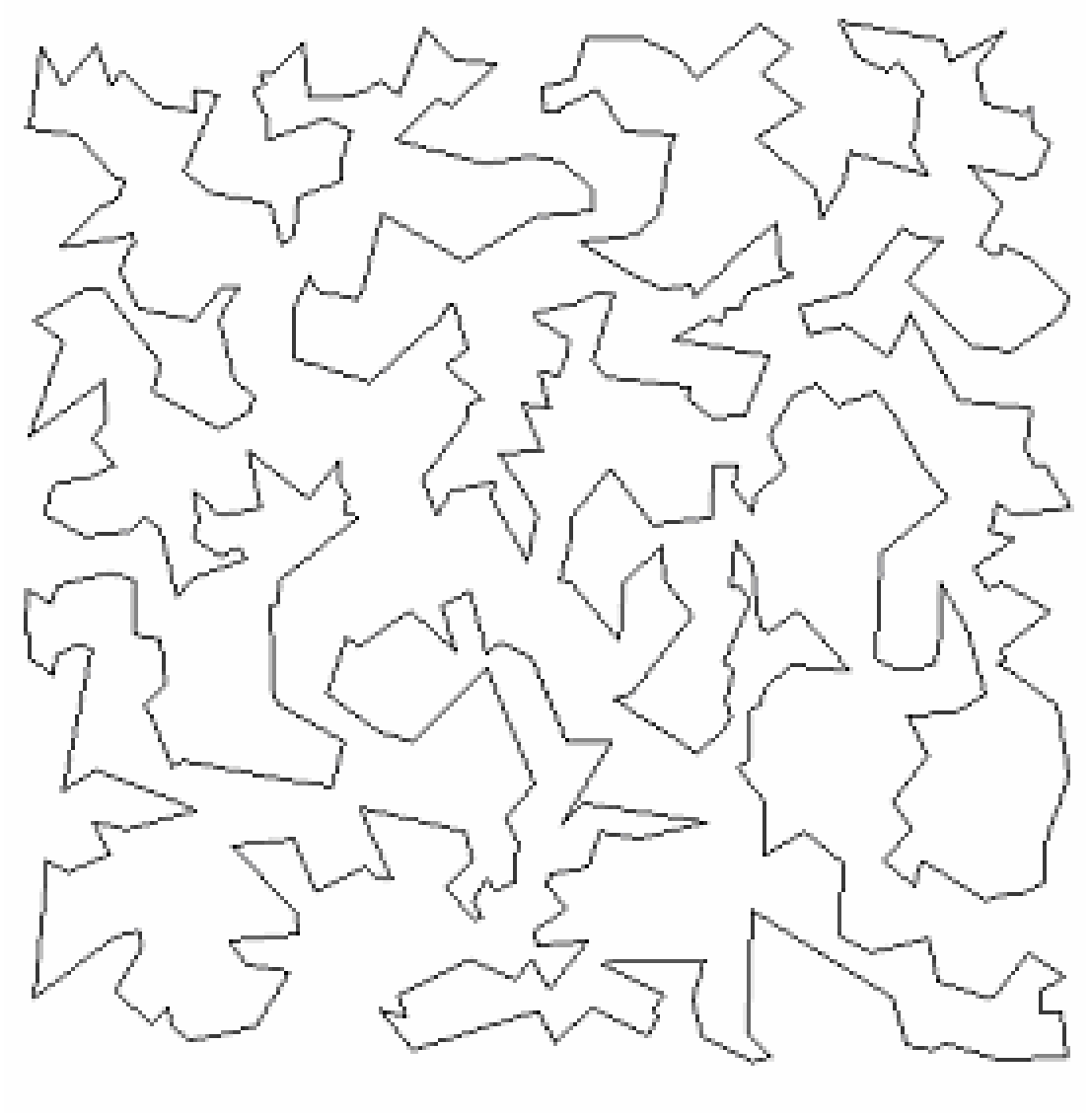

**Fig. 1 TSP solution for *N*=500 (computing time ~ 4 sec).**

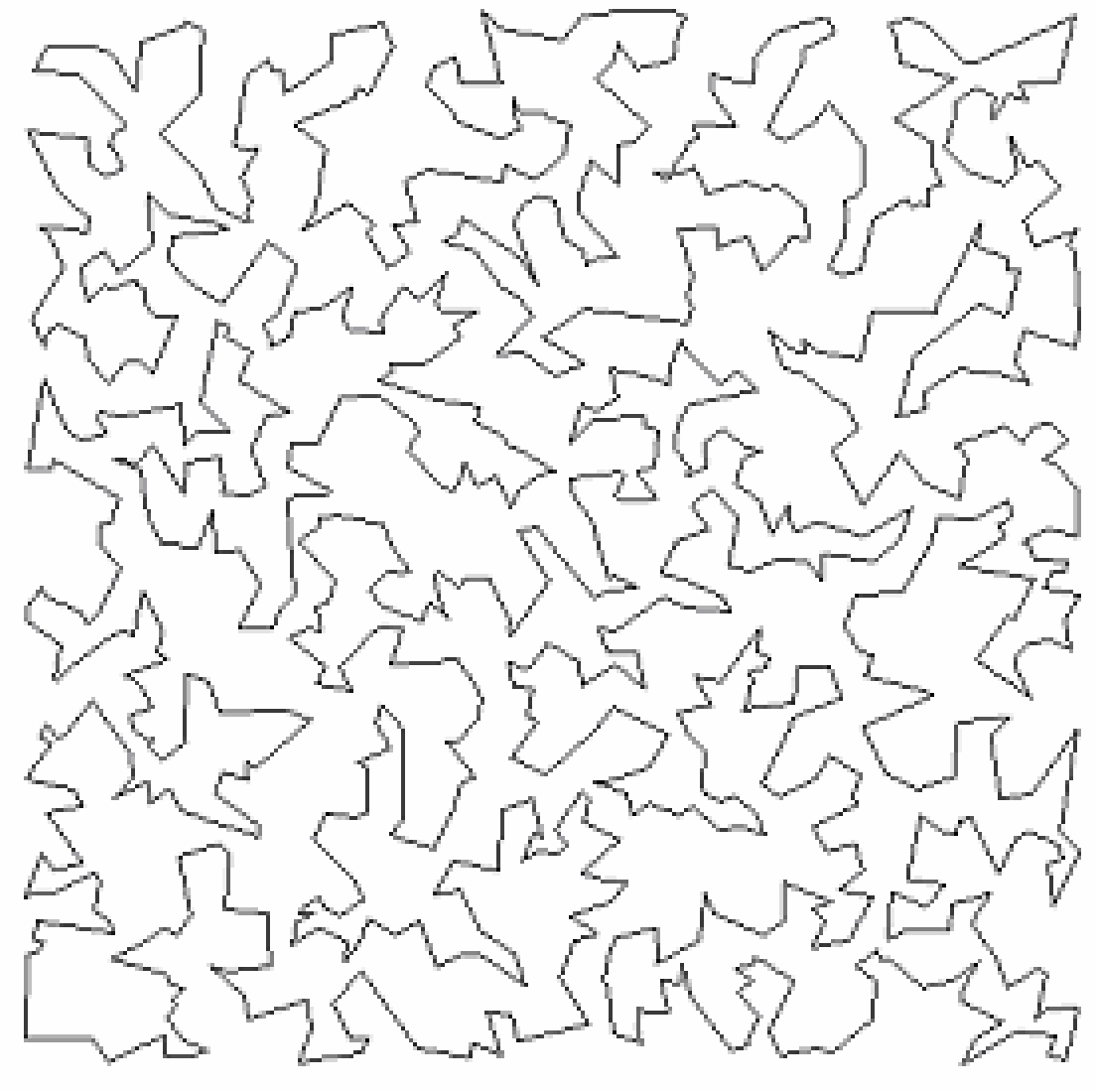

**Fig. 2 TSP solution for *N*=1000 (computing time ~ 38 sec).**

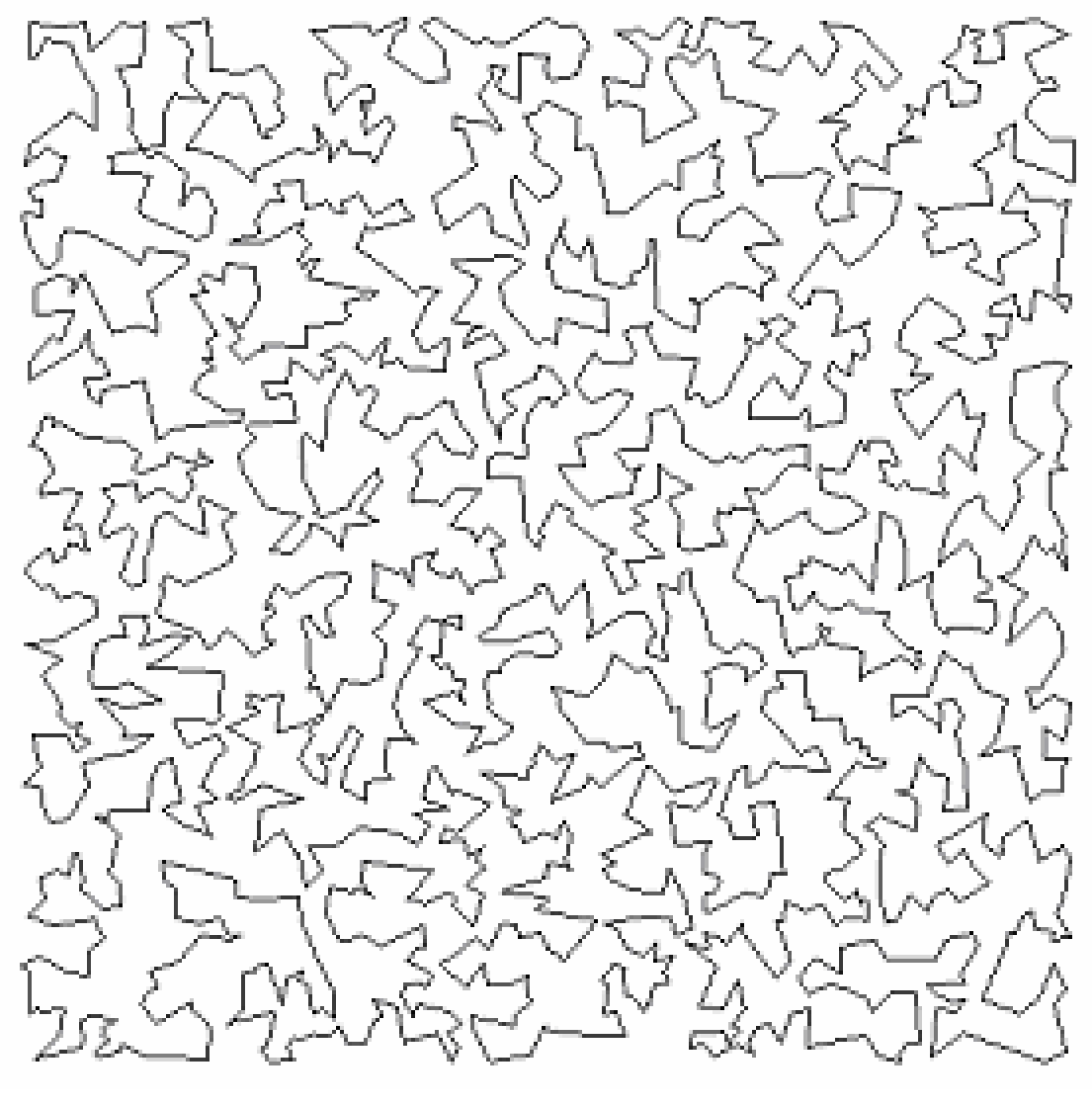

**Fig. 3 TSP solution for *N*=2000 (computing time ~ 431 sec).**

Computing times were ~ 4 sec for $N$=500 option; ~ 40 sec for $N$=1000 and ~ 400 sec for $N$=2000, i.e. doubling the $P^N$ volume leads to ten times increase in computing time. This provides the following estimate for real computational complexity of proposed method $K \sim O(N^{3.322})$; it may be deduced that this estimate and theoretical one demonstrate rather a good coincidence taking into account additional expenses relevant to operating system and compiler application.

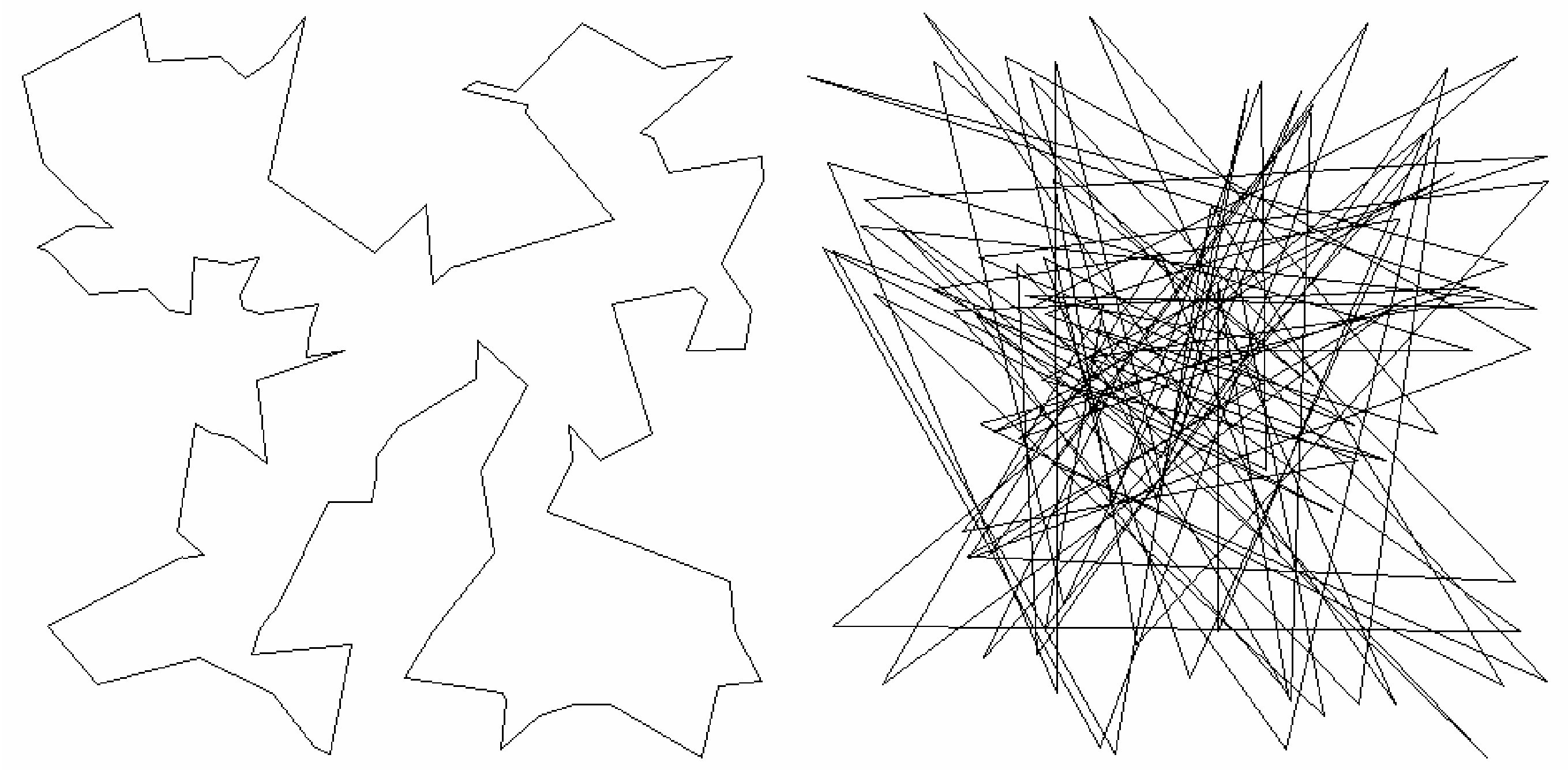

*a)* $D(\overline{R}^{i+1}) = D(\overline{R}^{i}) + \max_{r_k \in \overline{R}^i} \min_{p_j \in P^N - \overline{R}^i} \Delta(r_k, p_j)$ *b)* $D(\overline{R}^{i+1}) = D(\overline{R}^{i}) + \min_{r_k \in \overline{R}^i} \max_{p_j \in P^N - \overline{R}^i} \Delta(r_k, p_j)$

**Fig. 4 Variants of TSP solution for N=100 case.**

In closing it may be said that $D(\overline{R}^{i+1}) = D(\overline{R}^{i}) + \min_{r_k \in \overline{R}^i} \max_{p_j \in P^N - \overline{R}^i} \Delta(r_k, p_j)$ procedure will provide the $\overline{R}^N$ optimal route, which meets $D(\overline{R}^N) = \sum_{i=1}^{N, r_{N+1}=r_1} d(r_i, r_{i+1}) \to \max$ condition (this procedure to be initialized via computing $\overline{R}^2 : D(\overline{R}^2) \to \min$ and $\overline{R}^3 : D(\overline{R}^3) \to \min$ optimal sub-routes). Fig. 4 presents the computed optimal tours in both ***maxmin*** and ***minmax*** cases for the same $P^{100}$ points set.